\documentclass[12pt]{article}

          \usepackage{amsfonts}
          \usepackage{amssymb}

\def\be{\begin{eqnarray}}
\def\ee{\end{eqnarray}}
\def\bee{\begin{eqnarray*}}
\def\eee{\end{eqnarray*}}

\newtheorem{thm}{Theorem}

\newtheorem{lemma}[thm]{Lemma}

          \def\QED{{\bf QED}}

          \def\tr{\hbox{Tr}}

\def\ot{\otimes}

\def\Tr{{\rm Tr}}

            \title{
Maximal p-norms of 
entanglement breaking channels}
\author{Christopher King
\\ Department of Mathematics
\\ Northeastern University
\\ Boston MA 02115
\\
{\normalsize king@neu.edu}
}
\begin{document}

\maketitle

\begin{abstract}
It shown that when one of the components of a product channel is
entanglement breaking, the output state with maximal p-norm is always a product state.
This result complements Shor's theorem that both minimal entropy and 
Holevo capacity are additive for entanglement breaking channels.
It is also shown how Shor's results can be recovered from the p-norm results
by considering their behavior for p close to one.
\end{abstract}

\pagebreak



Holevo \cite{Hol2} introduced the following class of channels:
\be \label{eq:holv}
  \Phi(\rho) = \sum_{k=1}^K \, R_k ~ \tr \, \Big( X_k \rho \Big)
\ee
where each $R_k$ is a density matrix and where the  $\{X_k\}$ form
a POVM, that is $X_k \geq 0$ and $\sum X_k = I$. 
As Shor pointed out \cite{Shor}, channels of this form are entanglement breaking,
meaning that the state $(\Phi \ot I) (\rho_{12})$ is separable
for any bipartite state $\rho_{12}$. For this reason these channels 
are now known as entanglement breaking (EB) channels.
Shor proved additivity of the minimal entropy and the
Holevo capacity for EB channels \cite{Shor}, thereby settling the
question of their classical information-carrying capacity.

The purpose of this note is to show that EB channels also satisfy another additivity--type
property involving the maximal $p$-norm.
This notion was introduced by Amosov, Holevo and Werner \cite{AHW},
and involves the following non-commutative version of the 
usual $l_p$ norm for $p \geq 1$:
\be
|| A ||_{p} = \Big( \tr \, |A|^p \, \Big)^{1/p}
\ee
The maximal $p$-norm of a channel $\Omega$ is defined to be
\be\label{def:nu}
\nu_{p}(\Omega) = \sup_{\rho} \, || \Omega(\rho) ||_{p}
\ee
where the $\sup$ runs over density matrices in the domain of $\Omega$.

\medskip
\begin{thm}\label{thm1}
Let $\Phi$ be an entanglement breaking channel, and let
$\Omega$ be an arbitrary channel. Then for any $p \geq 1$,
\be\label{AHW}
\nu_{p}(\Phi \otimes \Omega) = \nu_{p}(\Phi) \,
\nu_{p}(\Omega)
\ee
\end{thm}

\medskip
The proof of Theorem \ref{thm1} relies on an intermediate bound which
we state below as Lemma 2. To set up the notation, consider the action of
the channel (\ref{eq:holv}) on a bipartite state $\rho_{12}$:
\be\label{bip}
(\Phi \ot I) (\rho_{12}) = \sum_{k=1}^K \, R_k \, \ot \, {\tr}_{1}\, 
[  (X_k \ot I) (\rho_{12}) ]
\ee
Define
\be\label{def:x,G}
x_{k} & = & \tr \, [  (X_k \ot I) (\rho_{12}) ] \\
G_{k} & = & x_{k}^{-1} \, {\tr}_{1} \, [  (X_k \ot I) (\rho_{12}) ] \nonumber
\ee
Then (\ref{bip}) reads
\be\label{bip2}
(\Phi \ot I) (\rho_{12}) = \sum_{k=1}^K \, x_{k} R_k \, \ot G_k
\ee
where now $\{ R_k, G_k \}$ are all density matrices, and $x_k \geq 0$
with $\sum x_k = 1$. Also, writing $\rho_1 = {\tr}_{2} (\rho_{12})$ for the reduced density matrix
it follows from (\ref{bip2}) that
\be\label{Phi-red}
\Phi(\rho_1) =  \sum_{k=1}^K \, x_{k} R_k
\ee
Define the following $1 \times K$ block row vector:
\be\label{def:R}
R = \bigg(\matrix{(x_1 R_1)^{1/2} & \cdots & (x_K R_K)^{1/2}}\bigg)
\ee 
Then $R^{*}$ is a $K \times 1$ block column vector, and
(\ref{Phi-red}) can be rewritten as
\be
\Phi(\rho_1) = R \,\,R^{*}
\ee

\medskip
\begin{lemma}\label{lemma2}
For all $p \geq 1$,
\be\label{lem2}
{\rm Tr} \, \Big( (\Phi \ot I)(\rho_{12}) \Big)^{p} \leq
\sum_{k=1}^K \, {\rm Tr} \, [(R^{*} R)^p]_{kk} \,\,\, {\rm Tr} [(G_{k})^p]
\ee
where $[(R^{*} R)^p]_{kk}$ is the $k^{\rm th}$ diagonal block of the 
$K \times K$ block matrix $(R^{*} R)^p$.
\end{lemma}

\medskip
\noindent{\it Proof of Theorem \ref{thm1}}:
let $\rho_{12} = (I \ot \Omega)(\tau_{12})$ so that
\be\label{tau}
(\Phi \ot I) (\rho_{12}) = (\Phi \ot \Omega) (\tau_{12})
\ee
Then from (\ref{def:x,G}) it follows that
\be\label{eqn1}
G_k = \Omega \Big( x_k^{-1} \, {\tr}_{1}
[(X_{k} \ot I)(\tau_{12})] \Big) =
\Omega \Big( G_{k}' \Big)
\ee
where $G_{k}' = {\tr}_{1} [(X_{k} \ot I)(\tau_{12})]$ is a density matrix.
Therefore (\ref{def:nu}) implies that
\be
\tr [(G_{k})^p] \leq \nu_{p}(\Omega)^p
\ee 
Together with
(\ref{lem2}) and (\ref{tau}) this implies
\be
\tr \, \Big( (\Phi \ot \Omega) (\tau_{12}) \Big)^{p} & \leq & \nu_{p}(\Omega)^p \,
\sum_{k=1}^K \, \tr \, [(R^{*} R)^p]_{kk} \\ \nonumber
& = & \nu_{p}(\Omega)^p \, \tr [(R^{*} R)^p] \\ \nonumber
& = & \nu_{p}(\Omega)^p \, \tr [(R R^{*})^p] \\ \nonumber
& = & \nu_{p}(\Omega)^p \, \tr [\Phi(\rho_{1})^p] \\ \nonumber
\ee
where we used the fact that the matrices $R^{*} R$ and $R R^{*}$ share the same
nonzero spectrum (and where $\tr$ changes its meaning several times). 
Using again the definition of maximal $p$-norm
(\ref{def:nu}) we deduce
\be
\tr \, \Big( (\Phi \ot \Omega) (\tau_{12}) \Big)^{p}  \leq 
\nu_{p}(\Omega)^p \, \nu_{p}(\Phi)^p
\ee
Since this bound holds for all $\tau_{12}$ it follows that
\be
\nu_{p}(\Phi \ot \Omega) \leq \nu_{p}(\Phi) \, \nu_{p}(\Omega)
\ee
and this implies the Theorem since the right side of (\ref{AHW}) can be achieved with
a product state. \QED

\bigskip
\noindent{\it Proof of Lemma \ref{lemma2}}:
this is an application of the Lieb-Thirring inequality \cite{LT},
which states that for positive matrices $A$ and $B$, and any $p \geq 1$,
\be\label{L-T}
\Tr \bigg( A^{1/2} B A^{1/2} \bigg)^p \leq
\Tr \bigg( A^{p/2} B^p A^{p/2} \bigg)
= \Tr \bigg( A^p B^p \bigg)
\ee
If $B \geq 0$ and $C$ is a general (non-positive) matrix, then $C B C^{*}$ has the
same nonzero spectrum as the matrix $(C^{*} C)^{1/2} B (C^{*} C)^{1/2}$, so the
Lieb-Thirring inequality also implies that in this case
\be\label{L-T1}
\Tr \bigg( C B C^{*} \bigg)^p \leq
\Tr \bigg( (C^{*} C)^{p} B^p \bigg)
\ee
Recall (\ref{bip2}), and define
\be
F_k & = & (x_k R_k)^{1/2} \ot I \\
H_k & = & I \ot G_k
\ee
Then (\ref{bip2}) can be rewritten as
\be\label{bip3}
(\Phi \ot I) (\rho_{12}) = \pmatrix{F_1 & \cdots & F_K}
\, \pmatrix{H_1 & \cdots & 0\cr \vdots & \ddots & \vdots\cr 0 & \cdots & H_K}
\, \pmatrix{F_1 \cr \vdots \cr F_K}
= F \, H \, F^{*}
\ee
where $F$ is the $1 \times K$ block row vector indicated,
and $H$ is the $K \times K$ diagonal block matrix. Applying (\ref{L-T1}) gives
\be
\tr \Big( (\Phi \ot I) (\rho_{12}) \Big)^p \leq
\tr \bigg( (F^{*} F)^{p} H^p \bigg)
\ee
Comparing with (\ref{def:R}) shows that
\be
(F^{*} F)^{p} = (R^{*} R)^p \ot I, \quad\quad
H_{k}^p =  I \ot G_{k}^p
\ee
and hence the result follows. \QED

\bigskip
As a further comment we note that Shor's results about additivity of minimal
entropy and Holevo capacity \cite{Shor} for EB channels can also be 
derived easily from Lemma \ref{lemma2}.
Taking the derivative of (\ref{lem2}) at $p=1$ gives
\be\label{der}
S \Big( (\Phi \ot I) (\rho_{12}) \Big) \geq
S \Big( \Phi(\rho_1) \Big) + \sum_{k=1}^K x_k S( G_k )
\ee
Again letting $\rho_{12} = (I \ot \Omega)(\tau_{12})$ and using
(\ref{eqn1}) it follows that
\be
S( G_k ) = S \Big( \Omega( G_{k}' ) \Big)
\ee 
where $G_{k}'$ is a density matrix. Using the definition of minimal entropy
\be
S_{\rm min}(\Omega) = \inf_{\rho} S \Big( \Omega (\rho) \Big)
\ee
it follows from (\ref{der}) that
\be
S_{\rm min}(\Phi \ot \Omega) \geq 
S \Big( \Phi(\rho_1) \Big) + \sum_{k=1}^K x_k S_{\rm min}(\Omega)
\geq S_{\rm min}(\Phi) + S_{\rm min}(\Omega),
\ee
which immediately implies the additivity of $S_{\rm min}$.

The additivity of Holevo capacity also follows easily from
(\ref{der}). It is convenient to first introduce a new quantity,
the minimal average entropy of an output ensemble from the channel,
for a fixed average input state $\rho$:
\be
S_{\rm av}(\Omega; \rho) = \inf_{\{p_k, \rho_k\}} \,\, \bigg[
 \sum_k p_k S \Big( \Omega(\rho_k) \Big) \,\, : \,\,
\sum p_k \rho_k = \rho \bigg]
\ee
As Matsumoto et al point out \cite{MSW}, the Holevo capacity of a channel $\Omega$
can be expressed in terms of this average output entropy:
\be\label{MSW}
{\chi}^{*}(\Omega) = \sup_{\rho} \bigg[
S \Big( \Omega(\rho) \Big) - S_{\rm av}(\Omega; \rho) \bigg]
\ee

\medskip
\begin{lemma}\label{lemma3}
Let $\Phi$ be an entanglement breaking channel, and let
$\Omega$ be an arbitrary channel. Then for any bipartite state
$\tau_{12}$,
\be\label{lem3}
S_{\rm av}(\Phi \ot \Omega; \tau_{12}) \geq
S_{\rm av}(\Phi; \tau_{1}) + 
S_{\rm av}(\Omega; \tau_{2})
\ee
\end{lemma}

\medskip
Lemma \ref{lemma3} follows easily from (\ref{der}), by
taking the average input state to be  $\tau_{12} = \sum p_k {\tau_{12}}^{(k)}$ and
applying the bound to each term in the sum
\be
\sum_k p_k S \Big( (\Phi \ot \Omega)({\tau}_{12}^{(k)}) \Big)
\ee
Then combining (\ref{lem3}) and (\ref{MSW}) with the subadditivity bound
\be
S \Big( \Phi \ot \Omega(\tau_{12}) \Big) \leq
S \Big( \Phi(\tau_1) \Big) + S \Big( \Omega(\tau_2) \Big)
\ee
immediately implies that
\be
{\chi}^{*}(\Phi \ot \Omega) \leq {\chi}^{*}(\Phi) + 
{\chi}^{*}(\Omega),
\ee
which establishes the additivity result for ${\chi}^{*}$.

\bigskip
{\bf Acknowledgements}
This work was partially supported by
National Science Foundation Grant DMS--0101205. Part of this work was
completed at a workshop hosted by the Mathematical Sciences Research Institute,
and the author is grateful to the workshop organisers and the Institute
for the invitation to participate.

{~~}

\end{document}